\documentclass[twocolumn,aps,pre,showpacs,superscriptaddress]{revtex4-2}
\usepackage{graphicx} 
\usepackage{amsfonts}
\usepackage{amssymb}
\usepackage{amsmath}
\usepackage{xcolor}
\usepackage{hyperref}
\usepackage{ulem}

\begin{document}

\title{Accurate Thermophysical Properties of Water using Machine-Learned Potentials}

\author{Tobias Hilpert}
\affiliation{%
University of Vienna, Faculty of Physics, Kolingasse 14-16, A-1090 Vienna, Austria
}
\affiliation{%
Vienna Doctoral School in Physics, Kolingasse 14-16, A-1090 Vienna, Austria
}

\author{Georg Kresse}
\affiliation{%
University of Vienna, Faculty of Physics, Kolingasse 14-16, A-1090 Vienna, Austria
}
\affiliation{%
VASP Software GmbH, Berggasse 21/14, A-1090 Vienna, Austria
}
\date{\today}

\begin{abstract}

Simulating water from first principles remains a significant computational challenge due to the slow dynamics of the underlying system. Although machine-learned interatomic potentials (MLPs) can accelerate these simulations, they often fail to achieve the required level of accuracy for reliable uncertainty quantification. In this study, we use MACE — an equivariant graph neural network architecture that has been trained using an extensive RPBE-D3 database — to predict density isobars, diffusion constants, radial distribution functions, and melting points.
Although equivariant MACE models are computationally more expensive than simpler architectures, such as kernel-based potentials (KbPs), their significantly lower total energy errors allow for reliable thermodynamic reweighting with minimal bias. Our results are consistent with those of previous studies using KbPs; however, equivariant models can be validated against the ground-truth density functional theory (DFT) ensemble, providing a critical advantage. These findings establish equivariant MLPs as robust and reliable tools for investigating the thermophysical properties of water with DFT-level accuracy.
\end{abstract}
\maketitle

\section{Introduction}
Water is an important resource for life and a solvent in sustainable energy applications. Justifiably so, computational insights into the structure of water have received large interest and follow a long history of over 50 years\cite{barker_structure_1969, rahman_molecular_1971, stillinger_improved_1974}. Simulations of liquid phase properties of water, particularly below its melting point, are subject to sluggish dynamics requiring significant computational expenses to achieve well-converged statistics\cite{gallo_slow_1996, starr_fast_1999, jansson_hidden_2010, omranpour_perspective_2024}. A strong effort in the development of empirical force fields tuned to accurately reproduce these properties has led to a selection of performant potentials that accurately reproduce water's liquid state\cite{jorgensen_quantum_1981, berendsen_interaction_1981, reimers_intermolecular_1982, jorgensen_comparison_1983, berendsen_missing_1987, mahoney_five-site_2000, yu_development_2003, horn_development_2004, rick_reoptimization_2004, abascal_general_2005, abascal_potential_2005, fanourgakis_flexible_2006, dyer_site-renormalised_2009, molinero_water_2009, gonzalez_flexible_2011, tainter_robust_2011, kiss_systematic_2013, fuentes-azcatl_non-polarizable_2014, izadi_building_2014, wang_building_2014, gonzalez_comprehensive_2016, izadi_accuracy_2016, jiang_hydrogen-bonding_2016}. 

Similar strides have been taken to repeat these successes from first-principles. The structure of water has been investigated particularly thoroughly using methods based on density functional theory (DFT), often achieving at least qualitative agreement to experiment\cite{laasonen_water_1992, laasonen_ab_1993, xantheas_ab_1995, sprik_ab_1996, grossman_towards_2004, vandevondele_influence_2004, baer_re-examining_2011, lin_structure_2012, pinnick_predicting_2012, ceriotti_nuclear_2013, gaiduk_density_2015, miceli_isobaric_2015, zen_ab_2015, chen_ab_2017, ruiz_pestana_quest_2018, cheng_ab_2019, zheng_structural_2018}. Computationally, first-principle calculations come at a much higher expense, allowing to achieve converged statistics only for a subset of observables at low temperatures\cite{gillan_perspective_2016, ruiz_pestana_quest_2018}.

Recent work on machine-learned potentials (MLPs) has shown the promise of significantly accelerating materials simulations at little loss of accuracy\cite{behler_generalized_2007}. Simulations of water using MLPs not only leverage the increased efficiency for a better understanding of water\cite{tai_no_description_1997, morawietz_neural_2012, morawietz_how_2016, morawietz_interplay_2018, singraber_density_2018,  gartner_signatures_2020, piaggi_homogeneous_2022, gartner_liquid-liquid_2022, montero_de_hijes_kinetics_2023}, but also act as a benchmark for different MLP architectures or training techniques\cite{zaverkin_predicting_2022, daru_coupled_2022, fu_forces_2023, kovacs_evaluation_2023, montero_de_hijes_comparing_2024, montero_de_hijes_density_2024, stolte_random_2025}. Water has therefore been explored with a wide variety of MLPs, for example GAP\cite{bartok_machine-learning_2013},
HDPNNS\cite{singraber_density_2018, cheng_ab_2019, wohlfahrt_ab_2020}, 
Deep potentials\cite{gartner_signatures_2020, zhang_phase_2021, piaggi_homogeneous_2022, gartner_liquid-liquid_2022} Allegro\cite{maxson_transferable_2024}, and MACE\cite{kovacs_evaluation_2023, ferretti_accurate_2025}. The training databases for these potentials were generated using a similarly large spread of DFT functionals, including PBE\cite{torres_using_2021}, SCAN\cite{zhang_phase_2021, gartner_signatures_2020, gartner_liquid-liquid_2022}, revPBE\cite{cheng_ab_2019, ferretti_accurate_2025}, and RPBE\cite{morawietz_how_2016, singraber_density_2018, buckova_density_2025, montero_de_hijes_comparing_2024, montero_de_hijes_density_2024, stolte_random_2025}. This has resulted in a range of predictions of important quantities, such as densities, diffusion coefficients, and melting points, sometimes varying even starting from the same or similar DFT ground truth\cite{buckova_density_2025, montero_de_hijes_comparing_2024}. Differentiating between these predictions is not trivial, as the uncertainty of a machine learning approach is difficult to evaluate. Computing energy or force errors of some subset of structures does not generally suffice to guarantee accurate observables\cite{fu_forces_2023}.

Thermodynamic reweighting is a technique widely used in Monte Carlo simulations to reweight an observable measured in one ensemble to another at a different temperature or Hamiltonian \cite{shen_statistical_2008, klimm_direct_2011}. The approach readily extends to molecular dynamics simulations\cite{nieto-draghi_histogram_2005, rick_increasing_2006}, and can be employed to reweight an observable computed by an MLP back to the ground-truth DFT ensemble. The reweighted prediction is, in principle, equal to the DFT result, despite the uncertainties introduced by the machine learning method. While this approach requires some additional DFT calculations and therefore incurs some additional computational cost, it is still significantly cheaper than direct DFT simulations. However, reweighting can only be performed with statistical significance if the sample and target distributions overlap sufficiently. If this is not the case, the reweighted result will disproportionally depend on very few outliers, which in practice cannot be sampled enough to yield reproducible results. In the context of MLPs, this requires the overlap of the MLP and DFT Boltzmann distribution to be large, achieved only if the total energy errors of the MLP are adequately low.  

It has been demonstrated across many materials that equivariant graph neural network potentials achieve significantly lower fitting errors than MLPs like GAP, KbPs, and BPNNs\cite{batatia_mace_2022, musaelian_learning_2023}. While the increase in accuracy comes with a sizable increase in computational cost, MLPs based on equivariant GNNs still show a significant speed-up compared to DFT. The increased accuracy can be leveraged to accurately reweight measured quantities back to the DFT ground truth with statistical significance.

We employ this approach to differentiate between previous results obtained by training several MLPs on a database generated using the RPBE exchange correlation functional with a Grimme D3 dispersion correction. RPBE-D3 water has been shown to be among the DFT exchange correlation functionals to reproduce experimental results reasonably well and has been extensively studied, with several MLP predictions reported\cite{morawietz_how_2016, singraber_density_2018, montero_de_hijes_comparing_2024, montero_de_hijes_density_2024, buckova_density_2025}. The predictions differ substantially depending on the employed MLP or training database\cite{morawietz_how_2016, montero_de_hijes_comparing_2024, buckova_density_2025}. While trends reported match qualitatively in all cases, the absolute value of quantities like the density maximum of RPBE-D3 simulated water remains uncertain. 

We show that training equivariant MACE potentials leads to MLPs that are accurate enough to enable statistically significant thermodynamic reweighting of the predicted density. The accuracy of these potentials elucidated some previously unnoticed inadequacies in the training database that were remedied. We then calculated the density isobars, diffusion constants, radial distribution functions, and melting points with these MACE potentials. The results obtained here agree well with previously published values obtained using KbPs, but could be reweighted using thermodynamic perturbation theory, confirming their validity.

\section{Methods}
All DFT calculations were performed using VASP 6.4\cite{kresse_efficient_1996, kresse_efficiency_1996} using high accuracy projector augmented wave potentials (O\_h and H\_h)\cite{blochl_projector_1994, kresse_ultrasoft_1999}, the RPBE exchange-correlation functional\cite{hammer_improved_1999} including a Grimme D3 dispersion correction\cite{grimme_consistent_2010} without Becke-Johnson dampening (zero dampening). 
For training of the machine-learned potentials, we considered systems containing 64 and 63 molecules using cubic cells. Systems with 64 molecules are prone to crystallizing into a cubic diamond-like structure, which leads to undesirable finite-size effects. The Brillouin zone was sampled only at the $\Gamma$-point. This DFT set-up was chosen to be consistent with previous work\cite{montero_de_hijes_comparing_2024,montero_de_hijes_density_2024}. Here, all structures were evaluated using a plane wave energy cut-off of $2000~\mathrm{eV}$. MACE\cite{batatia_mace_2022, batatia_design_2025} models were trained with two different settings: Invariant MACE with L=0 (128$\times$0e), and equivariant with L=2 (128$\times$0e+128$\times$1o+128$\times$2e). All models were trained on energies, forces, and stresses. Any other hyperparameter was kept equal between them. The equivariant model has roughly twice the number of trainable parameters of the invariant one at $880.000$. The cut-off radius was set to $6.0$ {\AA}  to match the cut-off of the KbPs. Models were trained for 250 epochs, employing a stochastic weight adjustment to increase the weights of energy and stress terms in the loss function for the last 50 epochs. For all other parameters, we refer to our previous publication and the data availability statement\cite{montero_de_hijes_comparing_2024}.

The training database is based on the water structures generated using active learning of VASP kernel-based potentials (KbP) reported in Ref. \cite{montero_de_hijes_comparing_2024}, recalculated with the set-up described above. Because of the greater accuracy of MACE, two inadequacies became evident during fitting the equivariant MACE potentials. Force errors on water structures sampled from molecular dynamics (MD) simulations using MACE potentials showed a systematic increase compared to force errors of the original database. This likely implies that MACE yields slightly different structures that are not fully represented using the kernel-based potential and the on-the-fly training relying mostly on the kernel-based potential. Additional structures generated from the MACE ensembles were added to the database until no meaningful decrease of the errors was observed. Additionally, all MACE models showed a non-zero mean shift in their stress predictions with respect to DFT ground truth. Trial DFT calculations on a subset of the training database with an increased plane wave cut-off of $4000~\mathrm{eV}$ confirmed a residual Pulay stress of $50\pm16~\mathrm{bar}$. This mean stress shift was added a posteriori to the diagonals of the stress tensor for all training structures contained in the database. The final database contained $2370$ structures, each containing $63$ or $64$ water molecules, for a total of $452.127$ atom entities (and three times as many force components). KbP results reported in this work were retrained with the final database, but showed no significant differences from previous results based on the initial data only. This means that the inherent errors of the KbPs somewhat obviate the need for an extremely accurate training dataset, whereas MACE requires very clean data to fully exploit its potential.

Molecular dynamics (MD) simulations used a time step of $1~\mathrm{fs}$ and a hydrogen mass increased to $8~\mathrm{amu}$ to allow for an accurate integration of the equation of motion. Diffusion constants were calculated with a hydrogen mass of $1~\mathrm{amu}$, and a time step of $0.5~\mathrm{fs}$.
MD simulations with KbPs were run in VASP 6.4\cite{kresse_efficient_1996, kresse_efficiency_1996}, using a Langevin thermostat and a Parinello-Rahman barostat, with all friction constants set $2~\mathrm{ps}^{-1}$. NpT simulations used a fictitious mass of the lattice of $8000$.
All MACE simulations in cells containing more than $128$ molecules were performed in VASP using entry points to supply energies and forces from the MACE ASE calculator\cite{batatia_mace_2022, hjorth_larsen_atomic_2017} ran with CUDA equivariance acceleration. The thermostatting settings were identical to the ones used for KbPs.
Parallel tempering simulations with MACE models were performed in Lammps\cite{thompson_lammps_2022, johansson_lammps-kokkos_2025} for four linked replicas using a Noose-Hoover chains thermostat and an MTK barostat with chain lengths set to $10$. The friction constants for the thermo- and barostat were set to $0.01~\mathrm{ps}$ and $0.1~\mathrm{ps}$, respectively. 

The extreme accuracy of L=2 equivariant MACE potentials comes with a significant increase in computational cost compared to the KbP simulations. Even when running single-precision simulations on an RTX6000 using cuda equivariance acceleration, simulations with equivariant MACE models were approximately 4 times slower than KbP simulations performed using CPUs only, and required significantly more memory. Furthermore,  for the system sizes considered here, the VASP MLFF requires between eight and 16 cores to achieve the best performance (see Ref. \cite{montero_de_hijes_comparing_2024}). On massively parallel CPU machines with multiple sockets, this allows one to perform parallel tempering calculations very efficiently, typically allocating 8 cores to each image and performing calculations with, for example, 16 (32) images on a total of 128 (256) cores. This is not possible using MACE, since the available implementation requires one GPU for each image. Consequently, considering the resources available to us, replica exchange calculations were limited to a small number of images for MACE.

\section{Results}
\subsection{Model evaluation}
Compared to KbPs, both invariant and equivariant MACE models yielded significantly lower errors in energies, forces, and stresses for both training and test sets. The equivariant MACE models consistently achieve force root mean square errors of $5$ meV/{\AA} with energy errors below $0.1$ meV/atom when evaluated on test sets containing structures of the same size as the ones contained in the training database. This is an improvement of up to a factor of $5$ compared to the kernel-based approach. To confirm consistent results for systems of varying sizes, we report errors on a separate test set containing structures with $128$ water molecules. 

These results are somewhat intriguing.
Firstly, we observe that the force error for the KbP does not increase from the holdout testset with 63/64 molecules to 128 atom test set, and that the energy per atom error decreases slightly. As discussed previously\cite{montero_de_hijes_comparing_2024}, root mean square errors (RMSE) \textit{per atom} should decrease as the square root of the number of atoms if there are no long-range correlation effects relevant on the scale of the reported error bars. This means that the RMSE of the \textit{total energy} increases as the square root of the number of atoms, which in turn relates to classical thermodynamics that states that the variance grows linearly with system size in the absence of long-range correlations. However, intriguingly, the energy error for the invariant MACE (L=0) increases from 64 to 128 atoms, even surpassing the KbP approach. This implies that the invariant MACE has fitted some unphysical long-range effects to the available data.  For equivariant MACE (L = 2), we observe a significant improvement; however, there is still a slight increase in error when transitioning from the holdout test set with 63/64 molecules to the evaluation set with 128 molecules. We believe this is due to the long-range interaction between the molecules being mainly dipole-related. Since dipoles are vector-like entities, faithful reproduction requires at least L = 1 equivariant message passing.  Our original KbP does not attempt to fit any long-range interactions and, remarkably, this makes it more robust in energy predictions for larger systems, albeit at significantly larger base errors. The KbP approach by construction entirely neglects any electrostatic long-range effects. Equivariant MACE (L = 2) performs significantly better than KbP for both the holdout test set and the larger 128-atom test set, but equivariant MACE still potentially suffers from some inadequacies in extracting long-range electrostatic interactions. However, the increase in errors is so small that we cannot rule out that the error increase is related to the different k-point densities in the 63/64 and 128-molecule ensembles (in both cases, sampling was performed at the $\Gamma$-point only).

\begin{table}[]
    \caption{Error metrics for MACE potentials and KbPs. Errors are evaluated on a test set containing 128 molecules drawn from MD simulations of other KbPs. Errors in brackets are holdout test-set errors evaluated on structures containing 63 and 64 molecules. The holdout test-set structures were removed from the training database before training. }
    \label{tab:errors}    \centering
    \begin{tabular}{crr}
              &  force RMSE [eV/{\AA}]& energy RMSE [meV/{atom}] \\
         \hline
         MACE L=2 & 8.9~ (~4.9)  &  0.09 (0.07)  \\
         MACE L=0 & 18.6~ (10.0) &  0.3~ (0.1~)  \\
         KbP &  27~~~~ (26~~) & 0.2~ (0.3~) \\
         \hline
    \end{tabular}

\end{table}

\subsection{Isobar and Density maximum}
\begin{figure}
    \centering
    \includegraphics[width=1.0\linewidth]{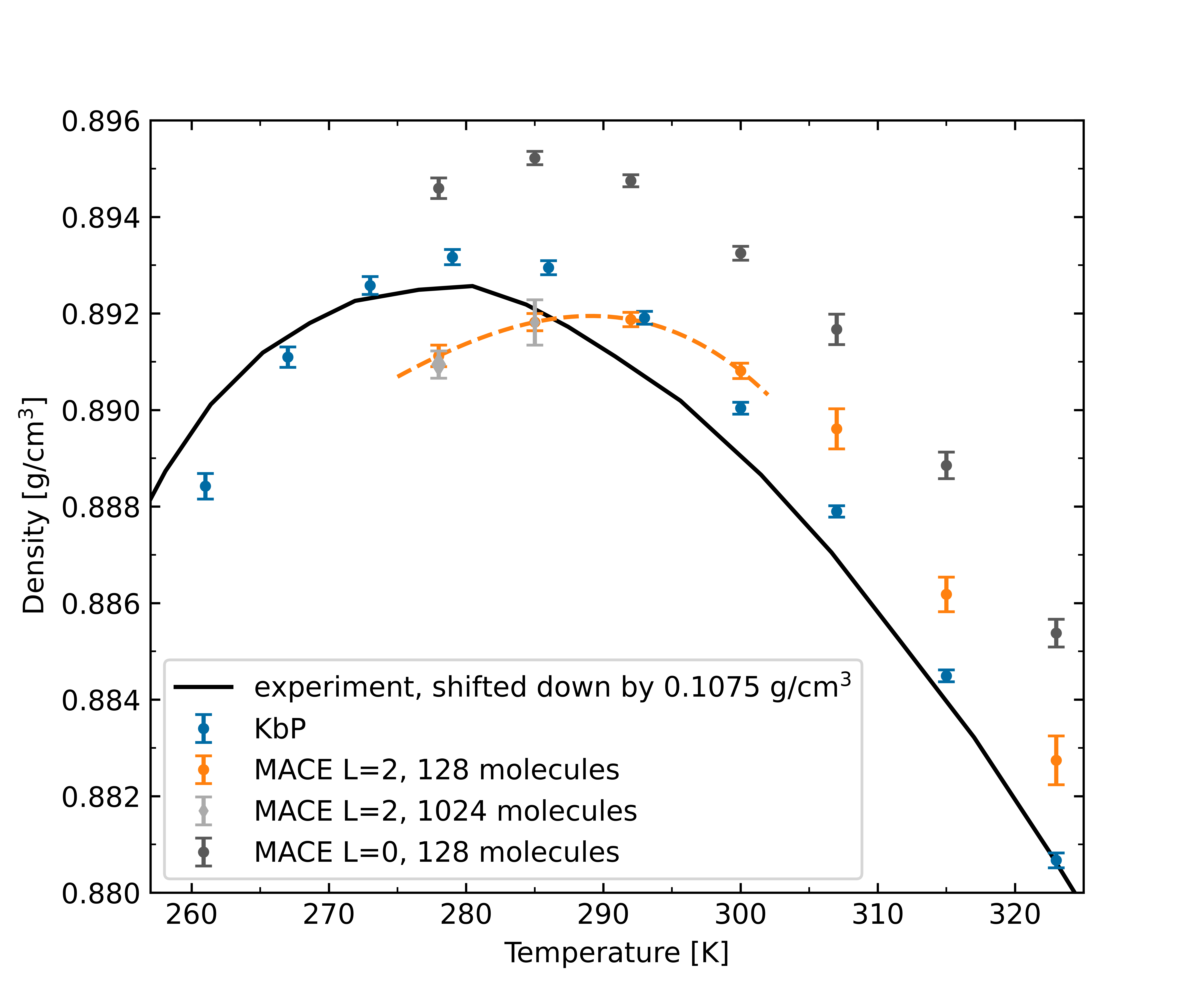}
    \caption{Density isobars of water for 128 molecule simulations obtained by parallel tempering for kbP and MACE. The data points for 1024 molecules were obtained without parallel tempering, but align within uncertainties with the 128 molecule results.}
    \label{fig:isobar}
\end{figure}

Density isobars were obtained from parallel tempering simulations with 128 water molecules for at least $10~\mathrm{ns}$. The results are shown in Fig. \ref{fig:isobar}, with error bars determined by block averaging with a block size of $500$. Equivariant MACE models and KbPs show reasonable agreement, particularly at temperatures above the density maximum. The MACE potential predicts a slightly lower density and a slightly greater temperature for the density maximum, but the curves look remarkably similar. Overall agreement between MACE and the KbP and our previous work\cite{montero_de_hijes_comparing_2024} is very good. The $128$ molecule cells have an average box dimension of $14$ {\AA}, preventing finite size effects for KbP simulations. The message passing formalism of 2-layer MACE models allows for information separated as far as $24$ {\AA} to still be directly transmitted. To rule out residual finite-size effects due to interactions between repeated images, separate simulations including $1024$ molecules were performed at two densities close to the density maximum. At the density maximum, the average length of the lattice vectors in these systems is above $30$ {\AA}. Because of the large memory requirement of simulating such large systems, we were not able to perform parallel tempering simulations for $1024$ molecules using the resources available to us. However, the results obtained show no indication of a finite-size effect.

Despite lower training and test set errors (with the previously noted caveat of larger errors for the total energy for 128 molecules), the results obtained from invariant MACE models appear to be further off from the reference equivariant MACE than the KbP. The shape of the simulated isobar shows a greater deviation from the experimental results than the other potentials. As we established in our tests on 128 molecules, total energy errors are largest for MACE L=0, so we excluded this potential from our further analysis. 

Finally, to control for possible noise introduced by the statistical optimizer, two MACE equivariant models were trained on the final database. The predicted volumes at $325~\mathrm{K}$ agreed to a difference below 0.1\%.

The density maximum of RPBE-D3 water fitted with equivariant MACE models was determined by fitting the isobar with a fourth-order polynomial suggests that the maximum lies at $288.8\pm2.4~\mathrm{K}$. This is in good agreement with the value determined by the kernels, which was $280~\mathrm{K}$ and $281~\mathrm{K}$ when calculated based on the old and new database, respectively. 

\subsection{Melting points}
\begin{figure}
    \centering
    \includegraphics[width=1.0\linewidth]{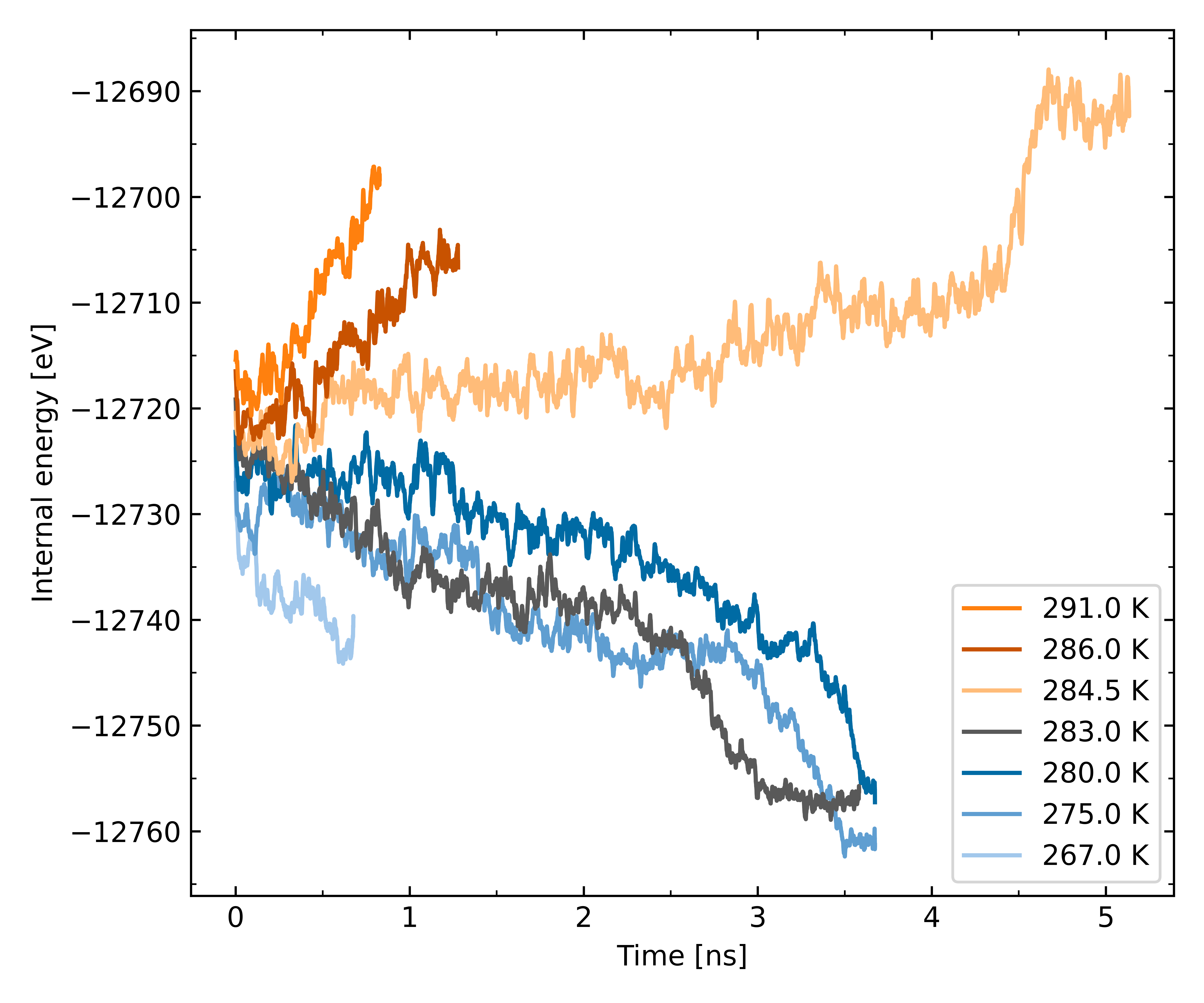}
    \caption{Internal energies of the interface simulations during melting and freezing using MACE L=2 potentials.}
    \label{fig:melting}
\end{figure}

Melting points were determined by simulating a cell containing a solid-liquid interface at several temperatures. To adequately represent both phases, the simulation cell contained a total of 800 molecules. During the simulations, the internal energies were monitored. A significant increase in the internal energies indicates that the solid part of the cell is melting, a decrease that the liquid is freezing. We confirmed this also by visually inspecting the final structures. The resulting internal energies are displayed in figure \ref{fig:melting}. The melting point was determined to be $283.75 \pm 1.0~\mathrm{K}$. This is $4~\mathrm{K}$ below the melting point determined by KbPs, but consistent with the higher temperature of maximum density determined from the isobar. The temperature difference between the density maximum and the melting point is $6~\mathrm{K}$, while it was $4~\mathrm{K}$ when simulated with the kernel-based approach. This is in good agreement with the experimental value of $4~\mathrm{K}$ and confirms that RPBE-D3 water reproduces the correct anomalous melting point at temperatures below the density maximum. We even speculate that improved statistics for MACE would bring the temperature difference in closer agreement with the KbP.

\subsection{Diffusion constants and RDF}
\begin{figure}
    \centering
    \includegraphics[width=1.0\linewidth]{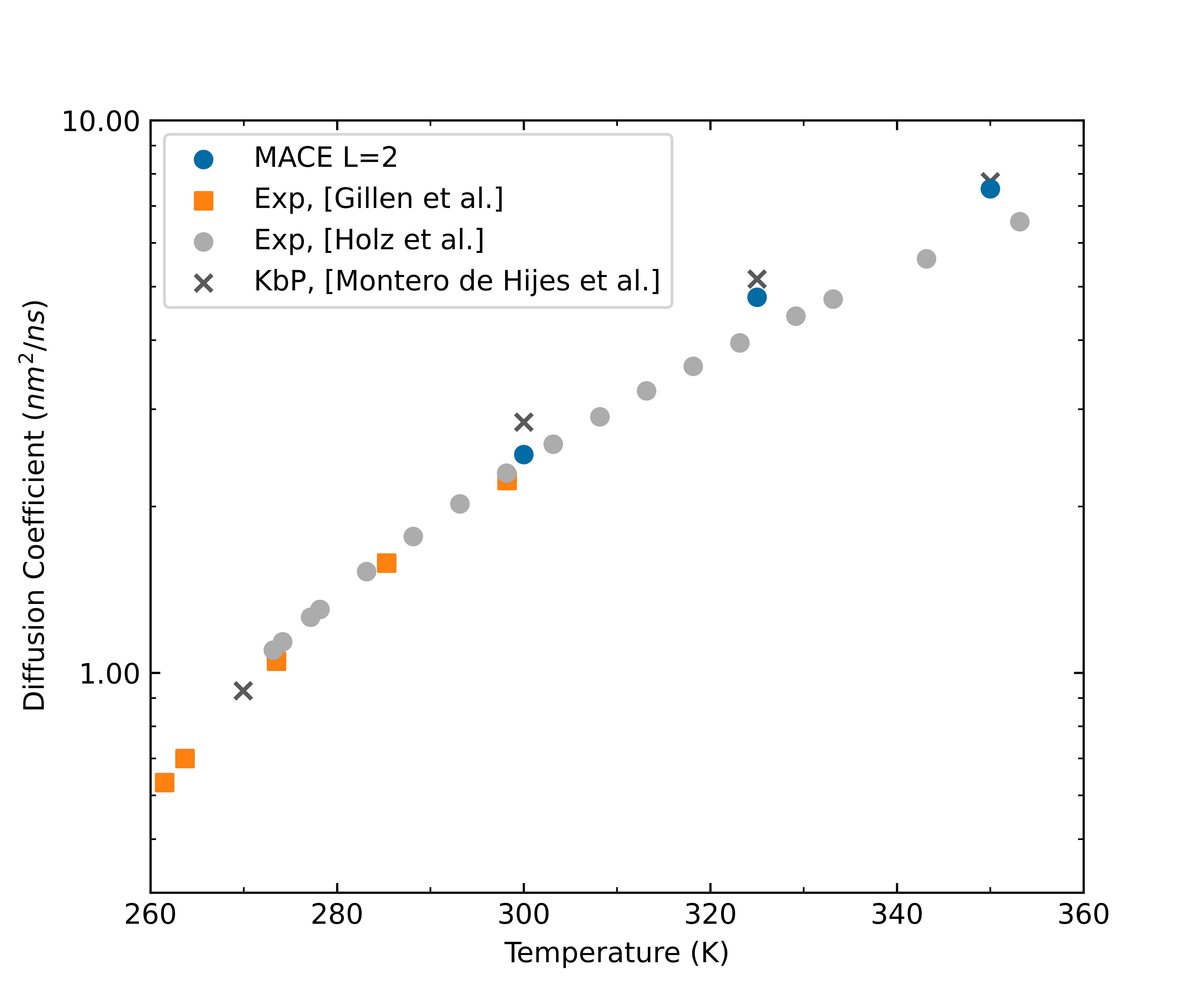}
    \caption{Diffusion constants obtained from the MACE L=2 potential. KbP results from \cite{montero_de_hijes_comparing_2024}, Experimental data reproduced from \cite{gillen_selfdiffusion_1972, holz_temperature-dependent_2000}.}
    \label{fig:diffusion}
\end{figure}

To calculate the diffusion constants, MD simulations were equilibrated in an NpT ensemble for each temperature with a decreased timestep of $0.5~\mathrm{fs}$ and a hydrogen mass of $1$. Once the average volume ceased to drift, a structure was extracted from the trajectory with a volume matching the average. The MD simulation was then continued from this starting point in an NVT ensemble. After equilibration, a structure was selected where the total energy equaled the average. The diffusion coefficients were determined from the mean square displacement of the NVE simulations starting from this structure. 
No significant differences between the MACE potentials and KbPs were observed. 

Radial distribution functions were determined from the same trajectories. Again, we show a comparison with the previous KbP results\cite{montero_de_hijes_comparing_2024}, showing only a minute differences. These are likely related to the different densities at which the simulations were performed. While the diffusion coefficients are in excellent agreement with experiment, the radial distribution functions show that both potentials are overstructured compared to the experiment. It is possible that part of this error can be attributed to the absence of nuclear quantum effects\cite{gillan_perspective_2016}: clearly, the classical simulations at T=325~K are somewhat in better agreement with experiment for the heights of the peaks (raising the temperature approximates the nuclear quantum effects to some extent).  Furthermore, the RDF peaks are shifted to larger volumes, which relates to the smaller observed densities. 

\begin{figure}
    \centering
    \includegraphics[width=1.0\linewidth]{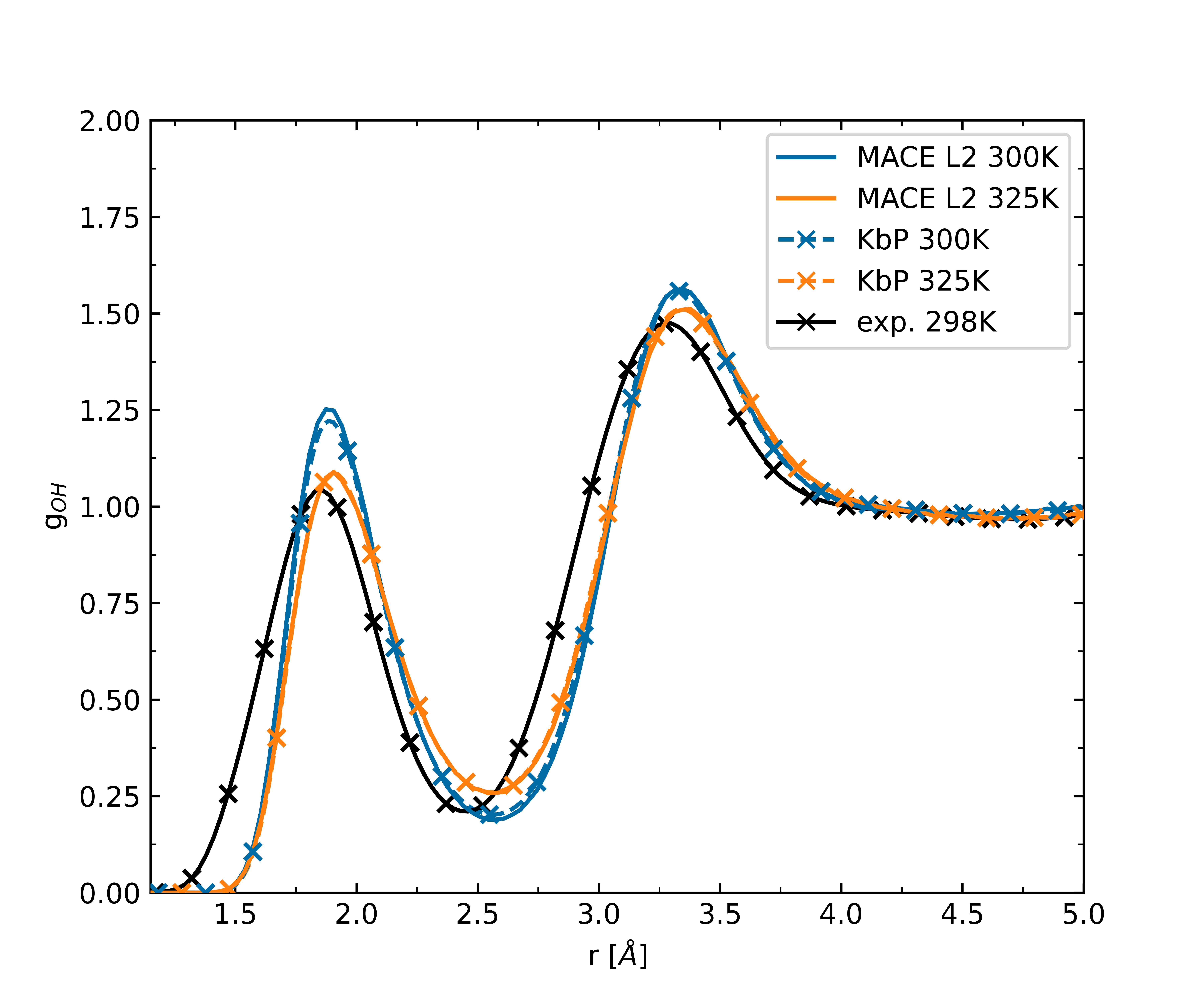}
    \caption{ Hydrogen-oxygen partial radial distribution function at different temperatures, calculated at the respective equilibrium volume with MACE L=2 potentials. The KbP results are from Ref. \cite{montero_de_hijes_comparing_2024}.}
    \label{fig:placeholder}
\end{figure}

\begin{figure}
    \centering
    \includegraphics[width=1.0\linewidth]{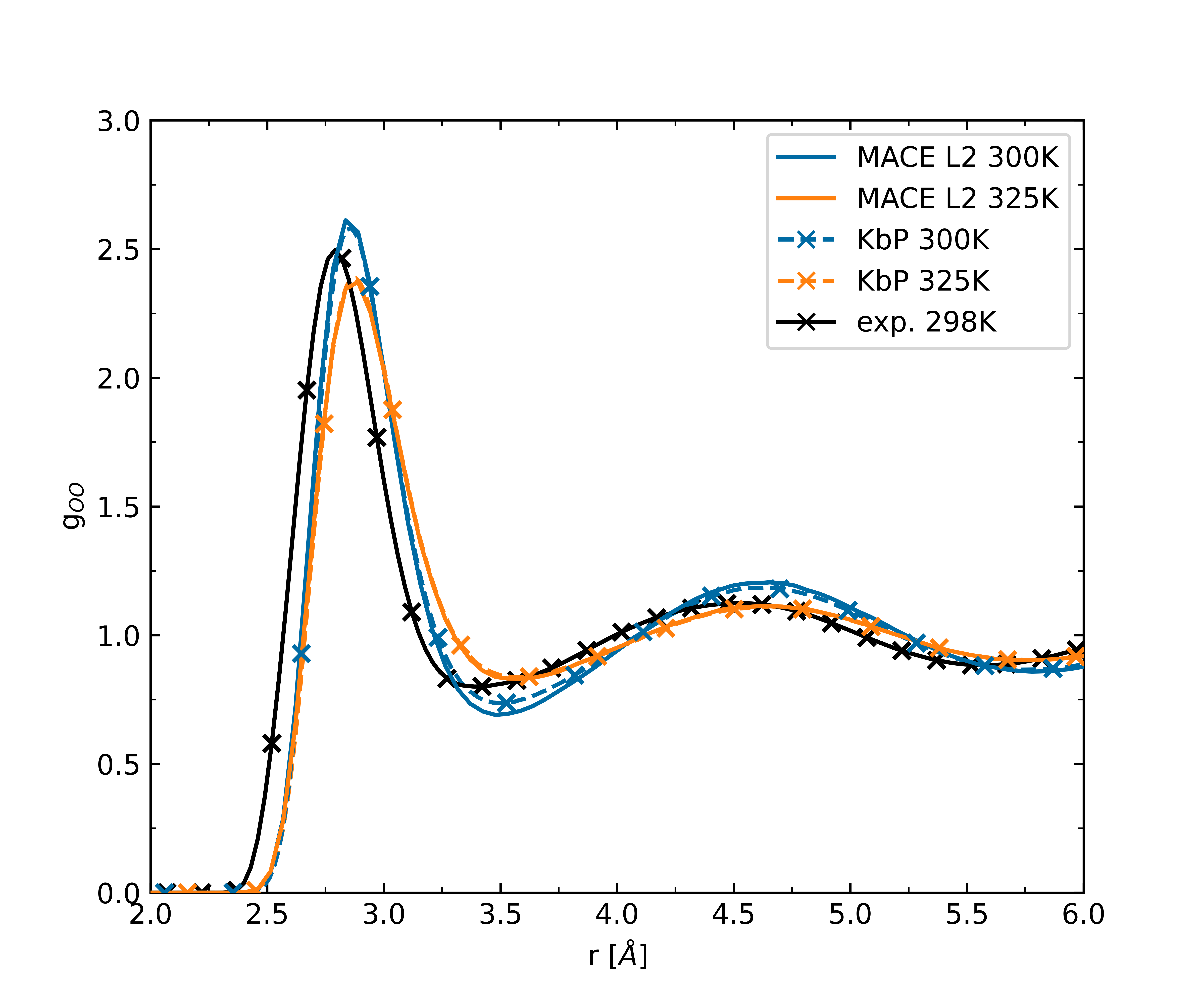}
    \caption{ Oxygen-oxygen partial radial distribution function at different temperatures, calculated at the respective equilibrium volume with MACE L=2 potentials.  The KbP results are from Ref. \cite{montero_de_hijes_comparing_2024}.}
    \label{fig:placeholder}
\end{figure}

\subsection{Thermodynamic reweighing}
\begin{figure}
    \centering
    \includegraphics[width=1\linewidth]{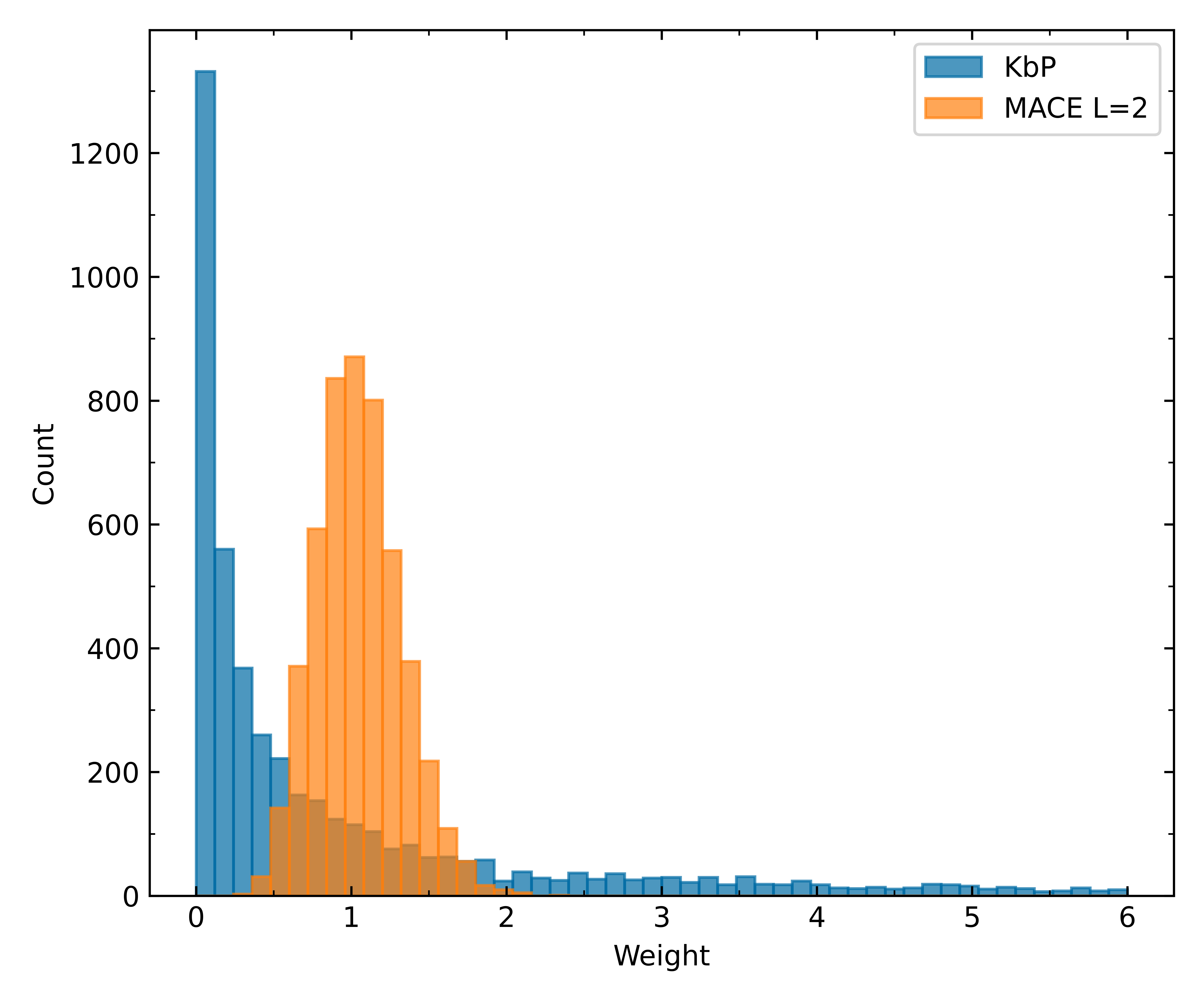}
    \caption{Weight distributions obtained for the KbP and MACE potential. The \textit{x}-axis is capped at $6.0$ for readability. Reweighing the KbP resulted in 520 structures with larger weights up to a maximum of 705.}
    \label{fig:hist}
\end{figure}
\begin{figure}
    \centering
    \includegraphics[width=1\linewidth]{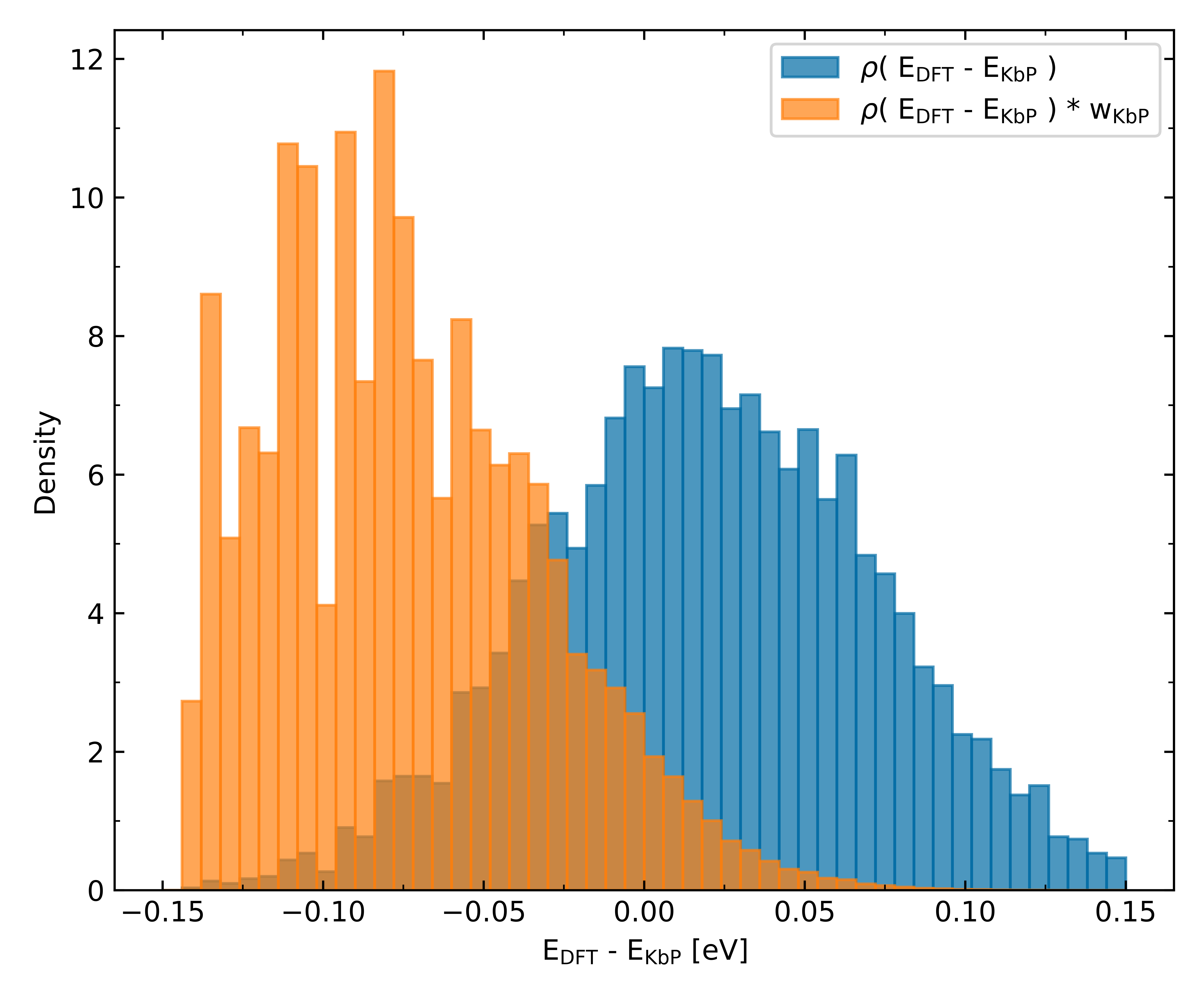}
    \caption{Distribution of energy differences between KbP and DFT for all sampled structures, along with the distribution multiplied by the weight $\exp \left[-\beta (\mathrm{E}_{\mathrm{DFT}}-\mathrm{E}_{\mathrm{KbP}}) \right]$.}
    \label{fig:kbp_rho}
\end{figure}

\begin{figure}
    \centering
    \includegraphics[width=1\linewidth]{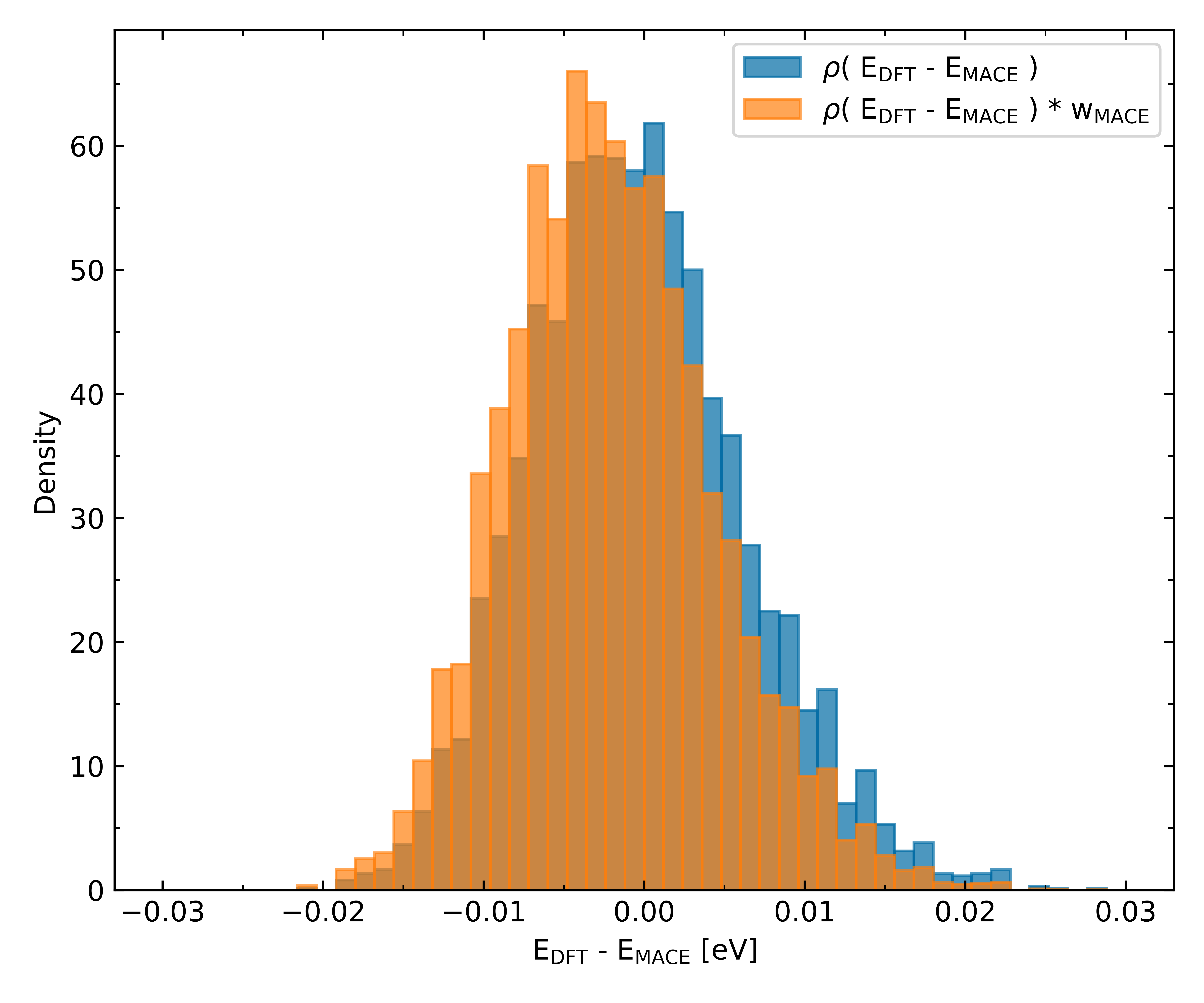}
    \caption{Distribution of energy differences between MACE and DFT for all sampled structures, along with the distribution multiplied by the weight $\exp \left[-\beta (\mathrm{E}_{\mathrm{DFT}}-\mathrm{E}_{\mathrm{MACE}}) \right]$.}
    \label{fig:mace_rho}
\end{figure}

In general, the MACE equivariant results agree to within 3\% with the results determined by KbPs, but show significant differences compared to other potentials like invariant MACE potentials and BPNNs\cite{montero_de_hijes_comparing_2024}. To differentiate between the results, we employ thermodynamic reweighting from the MLIP ensemble to the ground truth DFT ensemble. 
For one KbP and one equivariant MACE model, long simulations were performed in a cell containing 63 water molecules at $292~\mathrm{K}$. We have chosen this temperature because it is very close to the density maximum of the MACE model. The magnitude of the volume fluctuations was similar in both simulations, with a standard deviation in the volume of 2\%. To yield a low statistical standard error, 5000 largely uncorrelated structures were selected as a test set, ensuring that the average volume of the sample was similar to the total average over the simulation. Using 5000 structures,
we estimate that the error of the predicted mean volume 
should be below $2/\sqrt(5000)=0.03~\%$, if the reweighting is statistically reliable. This is more than sufficient for our present purpose.

Using the DFT energies, the average volumes of the selected sample were reweighted by the exponent of their energy difference between MLIP and DFT:

\begin{equation*}
\begin{split}
    \langle V \rangle_{\mathrm{DFT}} & \approx \frac{\sum_i^N  V (x_i) e^{-\beta [E_\mathrm{DFT} (x_i) - E_\mathrm{MLFF}(x_i)]}} {\sum_i^N e^{-\beta [E_\mathrm{DFT} (x_i) - E_\mathrm{MLFF}  (x_i)]}} \\
    & = \frac{\sum_i^N  V (x_i)w_i} {\sum_i^N w_i}
\end{split}
\end{equation*}
For the KbP, the large errors in total energies well above $25~\mathrm{meV}$ lead to poorly distributed weights, as shown in Fig. \ref{fig:hist} and Fig. \ref{fig:kbp_rho}. Specifically, in 
Fig. \ref{fig:kbp_rho} we see two issues. The reweighted density distribution clearly deviates from a Gaussian distribution on the left-hand side. We see that many configurations with an originally small weight (blue bar chart towards the left) are amplified in their weight, leading to a strongly frayed distribution at the left. 
Some of the light grey (orange) bars towards the left correspond to one or very few configurations. Furthermore, the volume varies greatly for a specific fixed $E_\mathrm{DFT} - E_\mathrm{MLFF}$, so that a clear covariance between $E_\mathrm{DFT} - E_\mathrm{MLFF}$ and the average volume is difficult to establish. This means that 
reweighting from KbPs to DFT shows extremely poor statistics because of the out-sized effect of very few structures in the sample. 
Reweighting from MACE to DFT (see Fig. \ref{fig:mace_rho}) does not reveal these effects. As the error metrics suggested (see Table \ref{tab:errors}), the distribution is about a factor of 5 narrower for MACE than for KbP. Specifically, the reweighted distribution on the left-hand side of the diagram retains a Gaussian shape, with frays that are no larger than those of the original distribution.

The effective sample size $N_\mathrm{eff}$ quantifies the statistical quality of the average:
\begin{equation*}
N_\mathrm{eff} = \frac{\left( \sum_i^N w_i \right)^2}{\sum_i^N w_i^2}
\end{equation*}
\begin{table*}[]
    \caption{Results for the density $\rho$ in g/cm$^3$ and volume $V$ in \AA$^3$. Results for the ensemble prepared using the machine-learned potential (MACE and KbP) and reweighted ensembles are shown at 292~K. Root-mean-square errors (RMSE) for the difference between the machine-learned energy and DFT energy are also shown.}
    \label{tab:thermoint}
    \centering
    \begin{tabular}{ccccc}
           & $\langle \rho \rangle$ ($V$) & $\langle \rho \rangle_{\rm reweighted}$ ($V$) & $\langle \rho \rangle_{\rm reweighted}$ ($V$)$^\dagger$ & RMSE  [meV] \\
         \hline 
         MACE L=2 & 0.8906 (33.59) & 0.8911 (33.57) & - & 6.75 \\
         KbP & 0.8903 (33.60) & 0.8893 (33.64) & 0.8905 (33.60) & 56.50 \\
     \hline
    \end{tabular}\\
    $^\dagger$ using capping (i.e. removing) outliers with an energy difference between KbP and DFT energies larger than 2.9~$\beta$.
\end{table*}
Compared to the parallel tempering simulations performed with 128 molecules, the 63-molecule simulations showed slightly lower densities for both potentials. The reweighted MACE result of 0.8911$~\mathrm{g}/\mathrm{cm}^3$ agrees perfectly with the result obtained from parallel tempering of 0.8914$~\mathrm{g}/\mathrm{cm}^3$, while the result obtained from KbPs suggests a lower density of 0.8893$~\mathrm{g}/\mathrm{cm}^3$ after reweighting. Reweighting the KbP ensemble shows poor statistics, however. The KbP sample achieved an effective sample size below $800$, raising the effective standard error of the reweighting. While it could be inferred that the resulting density is qualitatively correct, the large error prevents quantitative analysis. The equivariant MACE potential yielded well-distributed weights and an effective sample size of $4200$, confirming the exceedingly high accuracy of the reweighting in this case. The distributions of the resulting weights are displayed in Fig. \ref{fig:hist}
. "Capping" the weight distributions by excluding structures with out-sized influence can remedy poor effective sample sizes. As several structures in the KbP ensemble received a weight of over 100, capping can increase the effective sample size up to $1400$ when capping the distribution at $2.9 \beta$ and yielded a density of 0.8905 $~\mathrm{g}/\mathrm{cm}^3$, closer to the MACE results. However, this introduces additional uncertainty and bias if the reweighted average strongly depends on the capping criterion, as is the case for the KbP ensemble. Reweighting the equivariant MACE volumes did not alter the obtained results by more than 0.1\%, and capping the distribution at any point would decrease its effective sample size. Clearly, reweighing the MACE ensemble is possible with very high statistical significance, and equally important, the results did not change within the statistical error bars from the direct simulation results of MACE. This analysis suggests that the density isobars simulated with equivariant MACE potential are accurate to within at least 0.1\%. While the KbP results happen to match closely at this temperature, the reweighting and the frayed reweighted distribution lead to too noisy results to allow predictions with great statistical significance. Applying capping, the results for the KbP at T=292~K, however, are also consistent with the MACE results.

\section{Conclusion}

In this study, we used machine-learned interatomic potentials to determine highly precise thermophysical properties of RPBE-D3 water. We focused on the density isobars (including the density maximum), diffusion constants, radial distribution functions, and the melting point. Equivariant MACE models reproduce the trends previously reported for kernel-based potentials, but yield slightly higher temperatures for the density maximum ($289\pm3~\mathrm{K}$) and the melting point ($283.75\pm0.75~\mathrm{K}$) than a kernel-based approach. Diffusion constants remain unchanged and are in excellent agreement with experiment. The pair distribution function also remains in excellent agreement with previous calculations. Discrepancies observed in the radial distribution functions compared to experiments confirm that the simulated liquid is overstructured. This is consistent with the observed lower densities: at larger volumes the minima of the pair correlation function of water become more pronounced, whereas the position and height of the maxima hardly change. Another likely cause for the overstructuring is the neglect of nuclear quantum effects. A common method to approximate nuclear quantum effects is to increase the simulation temperature by about 10~\%, which indeed improves agreement with room temperature experiments in our case.

Crucially, the low total-energy errors of the equivariant MACE models enable thermodynamic reweighting of observables, such as the density, back to the DFT ensemble with statistical significance, providing a practical route to validate MLP predictions at the level of ensemble averages. The reweighting analysis confirms that the MACE-derived density at $292~\mathrm{K}$ is already within $\sim0.1$\% of the DFT result for the sampled conditions, whereas the substantially larger total-energy errors of kernel-based potentials lead to poorly behaved weights and prohibit quantitative validation without introducing additional bias through ad hoc procedures such as weight capping. These results clarify that reweighting is not a general remedy for inaccuracies in a potential; rather, it is only reliable when the MLP and DFT ensembles overlap sufficiently, which in practice requires very small total-energy errors. 

One of the most intriguing results of the present study is that invariant MACE produces very small fitting errors for systems containing a similar number of molecules to those in the training set. However, when applied to systems with twice as many atoms, invariant MACE performs poorly, with energy errors that even exceed those of the KbP. This suggests that message-passing networks are susceptible to fitting unphysical interactions, preventing accurate extrapolation of finite-size effects. A somewhat similar phenomenon occurs with equivariant MACE (L = 2), but in this case, the energy errors remain significantly lower than those of kernel-based methods. This implies that training datasets for equivariant message-passing networks require much greater care in their construction than for simpler, short-ranged machine-learned potentials. Specifically, we advise including systems of varying size to enable reliable extraction of long-range effects. We note that we did not attempt this in the present work, since to be entirely consistent, we would have had to perform many more calculations for 128 molecules and increase the k-point density for the 63 molecule calculations. Finally, MACE's supreme accuracy enabled us to identify slight inadequacies in our original training set. These were remedied by correcting the residual Pulay stress error — though small, this was still significant enough to degrade the fitting quality — and by adding additional training structures created using MACE.

\section*{Acknowledgments}
\noindent This research was funded in whole by the Austrian Science Fund (FWF) [10.55776/COE5] (Cluster of Excellence MECS). For open access purposes, the author has applied a CC BY public copyright license to any author-accepted manuscript version arising from this submission.

%
%
%

\section*{Data availablity}
The training and validation datasets produced in this work can be freely accessed at zendo (link to be added upon acceptance).


\bibliographystyle{apsrev4-2}
\bibliography{water_paper}

\end{document}